\documentclass{mn2e}
\usepackage{hyperref}
\usepackage{psfig}
\usepackage{epsfig}
\usepackage{lscape}
\usepackage{rotating}

\newif\ifAMStwofonts

\def\sqiglt{\hbox{\rlap{\lower.55ex \hbox {$\sim$}}\kern-.05em \raise.4ex \hbox{$<$}\,}}
\def\sqiggt{\hbox{\rlap{\lower.55ex \hbox {$\sim$}}\kern-.05em \raise.4ex \hbox{$>$}\,}}
\def\til{\ensuremath{\sim\,}}

\def\chisq{\ensuremath{\chi^2}}

\newcommand{\tim}[1]{\ensuremath{\times 10^{#1}}}
\def\deg{\ensuremath{^{\circ}}}
\def\etal{et al.\ }

\def\mwd{\ensuremath{M_{\rm WD}}}
\def\msol{\ensuremath{M_\odot}}
\def\xmm{\emph{XMM}}
\def\xmmn{\emph{XMM-Newton}}
\def\cms{\ensuremath{$cm$^{-2}}}
\def\cps{counts s$^{-1}$}

\def\xte{\emph{RXTE}}

\def\Mdot{\ensuremath{\dot{M}}}
\def\swift{\emph{Swift}}
\def\nh{\ensuremath{N_{\rm H}}}

\title[GK Per in outburst]{The unusual 2006 dwarf nova outburst of GK Perseii}

\author[Evans et al.]{P.A. Evans\thanks{pae9@star.le.ac.uk},
A.P. Beardmore , J.P. Osborne, G.A. Wynn
\\
Department of Physics and Astronomy, University of Leicester, Leicester, LE1 7RH, UK
}

\date{Accepted 
      Received }

\pagerange{\pageref{firstpage}--\pageref{lastpage}}
\pubyear{2007}

\begin{document}

\maketitle

\label{firstpage}

\begin{abstract} 
The 2006 outburst of GK Perseii differed significantly at optical and
ultraviolet wavelengths from typical outbursts of this object. We
present multi-wavelength (X-ray, UV and optical) \swift\ and AAVSO data,
giving unprecedented broad-band coverage of the outburst, allowing us to
follow the evolution of the longer-than-normal 2006 outburst across
these wavelengths. In the optical and UV we see a triple-peaked
morphology with maximum brightness \til1.5 magnitudes lower than in
previous years. In contrast, the peak hard X-ray flux is the same as in
previous outbursts. We resolve this dichotomy by demonstrating that the
hard X-ray flux only accounts for a small fraction of the total energy
liberated during accretion, and interpret the optical/UV outburst
profile as arising from a series of heating and cooling waves traversing
the disc, caused by its variable density profile.
\end{abstract}

\begin{keywords}
accretion, accretion discs -- novae, cataclysmic variables -- X-rays:
binaries -- stars:individual:GK Per
\end{keywords}

\section{Introduction}
\label{sec:intro}

\begin{figure*}
\begin{center}
\psfig{file=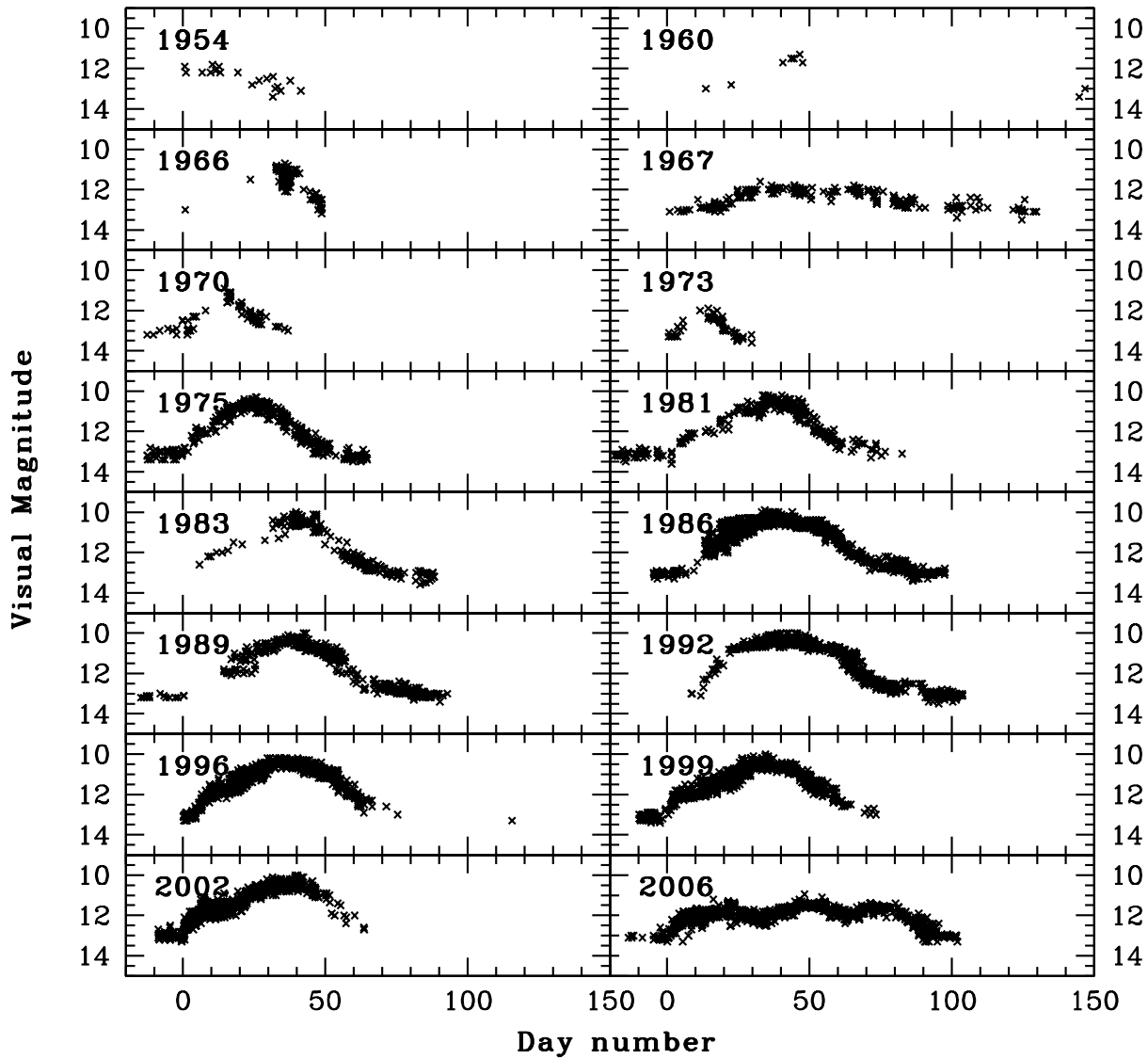,width=16cm}
\caption{AAVSO light curves of all outbursts of GK Per which can be
identified in the complete AAVSO dataset. Day zero is estimated by eye.}
\end{center}
\label{fig:allob}
\end{figure*}

The magnetic Cataclysmic Variable (mCV) star GK Per underwent an unusual
dwarf-nova-like outburst in 2006--2007. This system, which is not a
typical CV as it has a red-dwarf secondary and a 2-day orbital period
(Crampton, Cowley \&\ Fisher 1986; Morales-Rueda \etal2002), has been
observed to undergo  outbursts roughly every 3 years (e.g.\ Sabbadin \&
Bianchini 1983, Simon 2002). Its long orbital period and the fact that
it is an Intermediate Polar (IP, i.e.\ the white-dwarf primary has a
moderately strong magnetic field which truncates the inner accretion
disc; Watson, King \&\ Osborne 1985) make GK Per different from most
Dwarf Novae (DNe). The outbursts are still believed to be analogous to
normal DN outbursts, i.e.\ they are thought to be caused by enhanced
mass transfer through the accretion disc due to thermal instability
therein (e.g.\ Bianchini, Sabbadin \&\ Hamzaoglu 1982; Simon 2002,
Bianchini \etal2003).

GK Per has been well studied in quiescence and outburst. Its X-ray
emission is modulated at the 351-s white-dwarf rotational period (Watson
\etal1985) in quiescence and outburst. In outburst the
modulation is strong and single peaked, whereas in quiescence it is weak
and double peaked (Watson \etal1985; Norton, Watson \&\ King 1988;
Hellier, Harmer \&\ Beardmore 2004). Modulation at this period has also
been seen in optical spectroscopy (Morales-Rueda, Still \&\ Roche 1999)
and photometry (Patterson 1991). The American Association of Variable
Star Observers (AAVSO\footnote{http://www.aavso.org}) archive contains
data extending back to 1904, with frequent observations beginning in
1919. From 1954 onwards the light curve shows regular outbursts, peaking
typically around 10th magnitude. In Fig.~\ref{fig:allob} we show the
AAVSO light curve of every outburst  found in a visual inspection of the
dataset. As can be immediately seen, the 2006 outburst is fainter than
most, and shows an unusual morphology. It is, however, similar to the
1967 outburst.

The coverage of GK Per at X-ray and UV wavelengths is not as extensive
as in the optical, however it has been observed in quiescence and
outburst in both bands. Observations with \emph{Ginga\/} (Ishida \etal.
1992) and \emph{EXOSAT\/} (Norton \etal 1988; Watson \etal1985) show
that in hard X-rays (\til 2--10 keV) the typical outburst flux is \til10
times the quiescent flux. These, and RXTE observations (Hellier
\etal2004) show the typical outburst 2--10 keV flux to be
\til2.5\tim{-10} erg \cms s$^{-1}$. GK~Per is 470 pc away (McLaughlin
1960), thus $L_x\til6\tim{33}$ erg s$^{-1}$. \emph{IUE\/} observed GK
Per both in quiescence (Bianchini \&\ Sabbadin 1983) and outburst
(Rosino, Bianchini \&\ Rafanelli 1982), and saw a flux ratio of \til30
between the two observations at 2600~\AA.

The 2006 outburst of GK Per was announced by Brat \etal(2006) on 2006
December 18. Examination of the AAVSO light curve shows the outburst to
have begun on December 11; hereafter we use 2006 December 11 at 00:00 UT
(=JD 2454080.5, \swift\ MET 187488001.6 s) as the start time of the
outburst (hereafter `T0'). We obtained Target of Opportunity observations with
\swift\ (Gehrels \etal2004) which began on 2006 December 20 and were
repeated regularly throughout the outburst.

\section{Observations and data analysis}
\label{sec:obs}

\begin{table*}
\begin{center}
\begin{tabular}{cccccccc}
\hline
Obs Segment & Date and Time & XRT exposure & Mean XRT        & XRT spin  & UVOT exposure   & Mean UVOT & UVOT spin \\
            & start  (UT)   & (s)          & rate (s$^{-1}$)  & amplitude$^a$ & (s)             & rate (s$^{-1}$)  & amplitude$^b$ \\
\hline
001   &    2006-12-20 at 16:14  &  3946   & 1.58             & 39\%    &  3989           &   117           & 5.8\% \\
002   &    2006-12-26 at 02:29  &  4510   & 1.65             & 30\%    &  4539           &   155           & 4.7\% \\
003   &    2007-01-02 at 12:46  &  4697   & 1.98             & 21\%    &  4735           &   114           & 5.6\% \\
004   &    2007-01-09 at 08:35  &  4821   & 1.65             & 46\%    &  4838           &   99            & 7.9\% \\
005   &    2007-01-17 at 03:14  &  6017   & 2.02             & 27\%    &  5861           &   116           & 5.8\% \\
006   &    2007-01-23 at 02:12  &  5776   & 1.77             & 42\%    &  5798           &   198           & 5.6\% \\
007   &    2007-01-30 at 01:40  &  734    & 1.5              & 57\%    &  748            &   200           & 16.7\% \\
008   &    2007-02-04 at 14:52  &  3936   & 1.78             & 29\%    &  3978           &   197           & 3.8\% \\
009   &    2007-02-08 at 03:49  &  5250   & 2.24             & 42\%    &  5270           &   138           & 4.7\% \\
011   &    2007-02-12 at 02:20  &  2967   & 1.69             & 52\%    &  3023           &   113           & 11.2\% \\
012   &    2007-02-16 at 02:42  &  6155   & 2.06             & 35\%    &  6237           &   129           & 4.9\% \\
013   &    2007-02-19 at 14:30  &  2798   & 1.38             & 43\%    &  2808           &   206           & 9.6\% \\
014   &    2007-02-23 at 00:10  &  3239   & 1.48             & 40\%    &  3243           &   199           & 4.8\% \\
015   &    2007-02-26 at 00:29  &  6368   & 1.76             & 39\%    &  6418           &   188           & 3.5\% \\
016   &    2007-03-03 at 12:08  &  3144   & 1.46             & 47\%    &  3288           &   160           & 7.7\% \\
017   &    2007-03-06 at 04:42  &  7796   & 2.21             & 26\%    &  7848           &   86            & 5.2\% \\
018   &    2007-03-09 at 00:18  &  4201   & 1.73             & 25\%    &  4229           &   50            & 7.2\% \\
019   &    2007-03-13 at 00:32  &  5892   & 1.22             & 22\%    &  5981           &   35            & 4.4\% \\
\hline
020   &    2007-09-27 at 14:07  &  2473   & 0.12             & $<30$\%   & \\
\hline
\end{tabular}
\caption{Summary of the \swift\/ observations of GK Per. The spin amplitude is defined as $(max-min)/(max+min)$. Observation
020 was taken in quiescence, 6 months after the outburst finished, and has no UVOT data.
\newline $^a$ Typical uncertainties \til 3\%
\newline $^b$ Typical uncertainties \til 0.5\%
}
\label{tab:obs}
\end{center}
\end{table*}

For the first 6 weeks of the outburst \swift\ observed GK Per for 6 ks
once a week. Each observation was spread over three snapshots (one
snapshot per 96-min \swift\ orbit) as \swift\ is in a low-Earth orbit.
The X-ray telescope (XRT, Burrows \etal2005) was in its automatic state,
able to choose its operating mode for itself based on the source count
rate (Hill \etal2005); it remained in Photon Counting mode for every
observation. The UV/Optical telescope (UVOT, Roming \etal2004) was
operating in event mode. We requested the uvw1 filter (with a central
wavelength of 2600~\AA\ and FWHM of 693~\AA, Poole \etal2008), so that
our results would be comparable with the \xmmn\ observations taken
during the optical rise phase of the 2002 outburst (Vrielmann, Ness \&\
Schmitt 2005)\footnote{The \xmm\ observations included the Optical
Monitor (OM) using the uvw1 filter and the band-pass of the filter is
the same as for the \swift-UVOT, although the latter has \til10\%\ more
effective area in this filter than the \xmm-OM.}. Based on Vrielmann
\etal(2005) we anticipated a UVOT count-rate of \til30 \cps, well below
the level at which coincidence loss becomes an issue.

\swift\ data are available within a few hours of the observations taking
place, and it became immediately apparent that the UVOT count rate was
much higher than anticipated, and showed large variations. Three of the
first four observations had a coincidence-loss corrected count rate of
\til115 \cps, while one of them was at \til160 \cps. Within the
individual observations no variations of this magnitude were seen. To
better sample this variability,  we extended our observing campaign to
twice-weekly from 2007 January 30 (T0+50 days), the additional
observation being \til4 ks in duration each time. A summary of each
observation is given in Table~\ref{tab:obs}. By the time of the final
observation GK Per had almost returned to quiescence. Unfortunately, it
was not possible to continue observing with \swift\ after this point as
GK Per was within 45\deg\ of the Sun -- \swift's observing limit.

The data were analysed using the \swift\ software\footnote{part of the
lheasoft package: http://heasarc.gsfc.nasa.gov/lheasoft/}. The data
reduction was performed using version 28 of the \swift\ software. XRT
light curves and spectra of each observation were built using the
software presented by Evans \etal(2009). We created light curves with 30
s bins. UVOT light curves were built following the standard approach:
the {\sc attcorrjump} tool was used to correct the spacecraft attitude
file and {\sc coordinator} to create sky coordinates for each UVOT
event. The {\sc uvotscreen} task was called to remove bad events before
{\sc uvotevtlc} was used to produce a light curve with 5-s bins. This
task takes source and background regions, and performs background
subtraction and coincidence--loss correction. The final, calibrated
source brightness is provided as a count rate, magnitude and flux
density. Since the UVOT data were not astrometrically corrected, we
examined each observation individually and produced a unique source
region for each snapshot.

We barycentrically corrected the data, using a single barycentric
correction per observation segment, since this varied by \til1 s per
observation. We also folded each observation on the 351-s spin period,
using the same zero-point (in the barycentric frame) for each
observation. The spin-period modulation was clearly detected in X-rays
and the UV in all outburst observations; the pulse fraction is given in
Table~\ref{tab:obs}. 

\section{Results}

\label{sec:res}
\begin{figure}
\begin{center}
\psfig{file=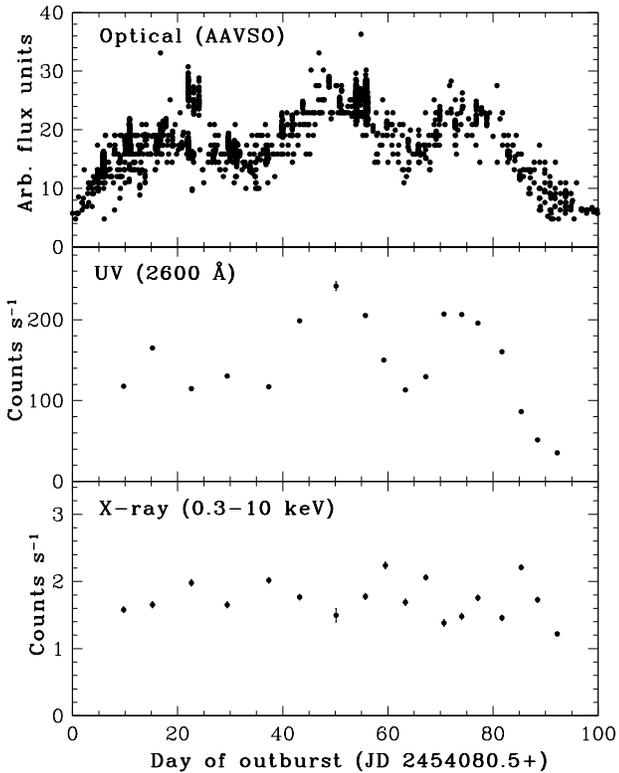,width=8.1cm}
\caption{The AAVSO, \swift-UV (2600 \AA) and \swift-X-ray (0.3--10 keV) light curves
of the 2006 outburst of GK Per. The AAVSO data have been converted to
(arbitrary) linear units for comparison with the \swift\ data. The
\swift\ data are binned to one point per observation.}
\label{fig:outburst}
\end{center}
\end{figure}

In Fig.~\ref{fig:outburst} we show the AAVSO optical light curve of the
2006 outburst with the X-ray and UV light curves  The optical light curve peaks
\til1.5 magnitudes fainter than most recent outbursts. The shape
of the light curve, rather than having a smooth `hump', shows a series of 3
humps. Note that the 2002 outburst showed a pause during the rise to
maximum and was very similar to the 2006 outburst for the first \til20
days, however thereafter the 2002 outburst returned to the `normal'
behaviour.

\begin{figure}
\begin{center}
\psfig{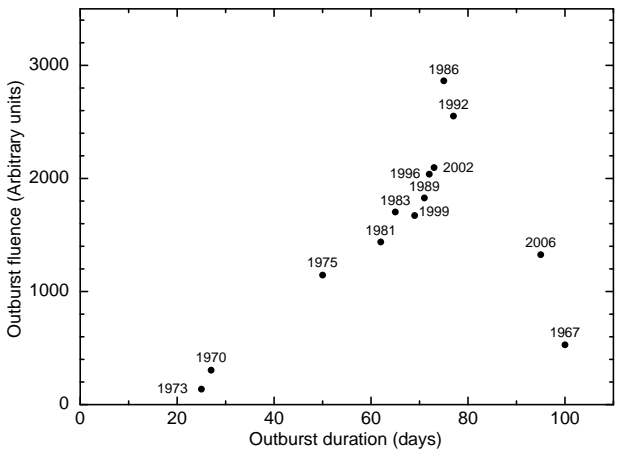}
\caption{Optical outburst fluence plotted against the duration.}
\label{fig:fluence}
\end{center}
\end{figure}

Since the 2006 and 1967 outbursts appear longer than the others, as well
as fainter, we measured the optical fluence of each outburst (i.e.\ the
flux integrated over the outburst). These are shown in
Fig.~\ref{fig:fluence}. Generally, as one would expect, outburst fluence
is correlated with duration. The exceptions are the 2006 and 1967
outbursts, which are long but of low fluence.

The UV light curve of the 2006 outburst is unique in its coverage and
thus cannot be compared to previous outbursts. However the ratio of the
maximum flux to that in the final observation (approximately quiescence;
the AAVSO magnitude was \til0.2 mag above the quiescent level) is
\til7.5 (=2.2 magnitudes) whereas the ratio of the \emph{IUE\/} flux at
2600~\AA\ between outburst and quiescence was \til28 (=3.6 magnitudes;
Bianchini \etal1983; Rosino \etal1982). The amplitude of the UV outburst
is thus around 1.5 magnitudes less than expected from previous data, as
has already been noted for the optical data. By analogy with the optical
data, we assume this implies a lower outburst flux rather than increased
quiescent flux. This is a surprising result, since the typical UVOT
count rate was nearly an order of magnitude higher than that reported by
Vrielmann \etal(2005). To investigate, we downloaded the pipeline
\xmm-OM products from the \xmm\ Science Archive and examined the uvw1
data (ObsID 0154550201). We found the count rate to be much higher than
claimed by Vrielmann \etal(2005), and comparable to or higher than in
our \swift\ data. Given that the \xmm\ data were taken on the rise of
the outburst, not the peak, we conclude that \xmm-OM data are consistent
with the idea that the UV emission in the 2006 outburst was fainter than
in previous outbursts. We also searched the OM data for evidence of
spin-period modulation, since this is clearly seen in our \swift-UVOT
data, but was reported as absent by Vrielmann \etal(2005). There is weak
evidence for spin-period modulation in the OM data, with an amplitude
\sqiglt3\%. The  presence of spin-period modulation in the UV emission
is thus not peculiar to the 2006 outburst.

While the UV light curve is clearly correlated with the optical one, the
X-ray light curve is not. By eye, some possible
anti-correlation or time-delayed correlation, with the UVOT data seems
possible. We thus performed a Discrete Correlation Function analysis
between the UVOT and XRT data, however no correlation was found above
the 1.8-$\sigma$ level. 

\begin{figure}
\begin{center}
\psfig{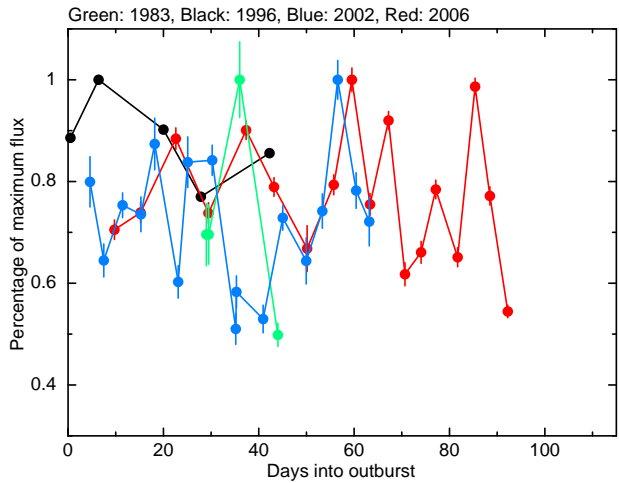}
\caption{A comparison of recent outbursts of GK Per in the X-rays. The
$y$-axis shows the count rate as a proportion of the maximum count-rate
observed (to normalise the different detectors). \emph{Green}:
2--10 keV \emph{EXOSAT} data from 1983 (Watson \etal1985). \emph{Black}:
2--15 keV \xte\ data from 1996 (Hellier \etal2004). \emph{Blue}: 2--15
keV \xte\ data from 2002. \emph{Red}: 0.3--10 keV \swift\ data (this
paper).}
\label{fig:xrayob}
\end{center}
\end{figure}

The 1983, 1996 and 2002 outbursts of GK Per were all monitored in the
X-rays with different satellites (Watson \etal1985; Hellier \etal2004).
The most extensive dataset prior to that presented here is unpublished
\emph{RXTE\/} monitoring of the 2002 outburst. In Fig.~\ref{fig:xrayob}
we show the X-ray flux evolution from these 3 outbursts in addition to
the \swift\ data from 2006. As can be seen, there is no systematic X-ray
evolution seen during outbursts, unlike in the optical. The 2006
outburst is however fairly typical in its relative flux evolution. The
2--10 keV flux during \swift\ observation 009 (the observation during
which the X-ray flux was greatest) was 3.3\tim{-10} erg \cms\ s$^{-1}$
($L_x=8.7\tim{33}$ erg s$^{-1}$), which is consistent with the peak hard
X-ray fluxes seen in previous outbursts (e.g. 1983, Watson \etal1985;
1989, Ishida \etal1992; 1996, Hellier \etal2004). In 2007 September
\swift-XRT observed GK Per in quiescence for calibration purposes
(unfortunately, the UVOT was not in operation). The 2--10 keV flux in
this observation was 2.3\tim{-11} erg \cms\ s$^{-1}$ ($L_x=6.1\tim{32}$
erg s$^{-1}$) i.e.\ a factor of 14 lower than in outburst. This is
similar to the 2--10 keV quiescent fluxes of 4.5\tim{-11} erg \cms\
s$^{-1}$ and 2.7\tim{-11} erg \cms\ s$^{-1}$ reported by Norton
\etal(1988). Thus, unlike the optical and UV, both the peak 2--10 keV
flux and the outburst/quiescence 2--10 keV flux ratio seen in this
outburst, are consistent with measurements from previous outbursts.

\subsection{X-ray Spectroscopy}
\label{sec:spec}

\begin{figure}
\begin{center}
\psfig{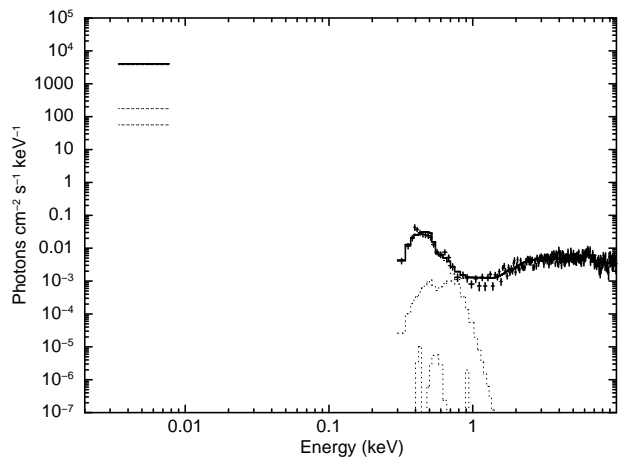}
\caption{The spectrum obtained from the observation 009 data, with the
best-fitting model applied. For fitting, the data were grouped
to contain at least one count per spectral bin; however for plotting
purposes the data have been binned such that each point is significant
at at least the 5-$\sigma$ level. The best-fitting model parameters are
detailed in Table~\ref{tab:spec}}
\label{fig:spec}
\end{center}
\end{figure}

We created X-ray spectra for each observation of GK Per and modelled
them in {\sc xspec 12.4}. We first used the {\sc grppha} tool to ensure
that there was at least one count per spectral bin, and performed
fitting using the $C$-statistic (this is more reliable than the \chisq\
statistic; see. e.g.\ Humphrey \etal2009).

The hard X-ray emission in IPs is believed to come from a dense,
post-shock plasma cooling via bremsstrahlung emission (e.g.\ Aizu 1973;
Cropper \etal1999), which we fitted with the physical model of this
developed by Cropper \etal(1999). A simple photoelectric absorber and
two partial covering absorbers were necessary to obtain a good fit to
the hard X-ray data, as previously found (e.g.\ Ishida \etal1992). 

There were still significant residuals seen at soft energies. A number
of IPs show evidence for a soft (\til30--100 eV) blackbody component in
their X-ray spectra (e.g.\ de~Martino \etal2004; Evans \&\ Hellier 2007;
Anzolin \etal2008), as did  GK Per during the 2002 outburst (Vrielmann
\etal2005; Evans \&\ Hellier 2007). We thus added a blackbody component.
We also added narrow Gaussian lines at 0.423, 0.557 and 0.907 keV to
reproduce the lines from the nova shell; the energies, widths and
normalisations of these lines were taken from Balman (2005).

In IPs it is often assumed that most of the accretion luminosity is
emitted as hard X-rays (e.g.\  Evans \&\ Hellier 2007 showed the
bolometric luminosity of the soft component to be \sqiglt0.1 of the
bolometric luminosity of the hard component), however since the 0.3--10
keV bandpass of the XRT (and the EPIC instruments on \emph{XMM}) covers
only the hard tail of the blackbody component, the details of the soft
emission are not particularly well constrained. To remedy this we
created a spectral point from the UVOT data for observation 009 (the
X-ray brightest observation) using the {\sc uvot2pha} tool and fitted
the combined UVOT and XRT data for this observation. The (unabsorbed)
bolometric flux from the blackbody component in this fit far exceeded
the hard X-ray flux; however at $D=470$ pc it also exceeded the
Eddington luminosity by more than an order of magnitude [assuming a 0.87
\msol\ white dwarf; Morales-Rueda \etal(2002)]. There thus cannot be a
single spectral component, powered by accretion energy, spanning the UV
to soft X-ray wavelength range.

A potential contributor to the UV emission is the inner disc. Frank,
King \&\ Raine (2002) give the temperature of the disc at radius $R$ as:

\begin{equation}
T(R)=\left( \frac{3GM\dot{M}}{8\pi R^3 \sigma}
\left[1-\left(\frac{R_{\rm
wd}}{R}\right)^{\frac{1}{2}}\right]\right)^\frac{1}{4}
\label{eq:tdisc}
\end{equation}

Using \mwd=0.87 \msol (Morales-Rueda \etal2002), the white dwarf
mass-radius relation of Nauenberg (1972), and assuming\footnote{by
taking $L_{x, 2-10}=GM\dot{M}/R_{\rm wd}$: this is a lower limit since
it assumes that all of the liberated accretion energy was radiated in
the 2--10 keV band)} $\dot{M} \ge 5\tim{16} $g s$^{-1}$, we find the
disc temperature at the corotation radius (=7\tim{9} cm) to be $T_{\rm
corot} \ge 12,400$ K, which corresponds to a blackbody peak wavelength
of $\lambda_{\rm corot} \le 2360$ \AA; towards the centre of the uvw2
filter bandpass. Thus we expect the inner disc to make a significant
contribution to the UVOT flux.

We therefore modified our model further, adding a second blackbody with
the peak wavelength fixed at 2360 \AA (with only a single UV spectral
point we cannot leave this parameter free). This blackbody was absorbed
only by a {\sc tbabs} component (which includes the effects of dust),
with \nh\ tied to that of the {\sc phabs} component acting on the harder
emission. The best-fitting spectrum is shown in Fig.~\ref{fig:spec} and
the parameters are tabulated in Table~\ref{tab:spec}.  In this fit, only
\til1\%\ of the combined flux of the harder blackbody and thermal plasma
components -- i.e.\ those expected to radiate the majority of the
liberated accretion energy -- is emitted in the 2--10 keV band,
suggesting that the hard X-ray flux is a poor proxy for accretion rate.
This 1\%\ figure should be seen as a poorly constrained lower limit to
due the limitations of this model fit: the harder blackbody component is
affected by the softer one, whose temperature is fixed at that
determined assuming that the 2--10 keV flux comprises 100\%\ of the
accretion flux, However, the fit shows that this is not the
case, i.e.\ the approach is not self-consistent. A self-consistent model
is not readily attainable however. In order to properly constrain the
spectrum and hence the wavelengths at which the accretion energy is
radiated we need simultaneous X-ray, and UV (preferably broad-band UV)
spectroscopy, which we do not have; nonetheless it is clear that a
significant portion of the accretion energy can be radiated below the
2--10 keV band.

\subsection{Spectral evolution through the outburst}

While some variation in best-fitting spectral parameters was seen
between observations, the uncertainties were too large to determine
whether there was any spectral evolution taking place during the
outburst. We tried combining several observations to give a total of 5
spectra for the outburst: `plateau 1' (observations 001--005), `hump 1'
(observations 006--008); `plateau 2' (009--012), `hump 2' (013--016) and
`fading' (017--018). The only parameter which showed significant
variation between these regions was the column density of the less dense
of the two partial covering absorbers, this was \til(8.4
$\pm$\til0.7)\tim{22}~\cms\ during the `humps' and \til(4.5
$\pm$\til0.5)\tim{22}~\cms\ during the `plateaux'.

\begin{table*}
\begin{center}
\begin{tabular}{cccl}
\hline
Component        &   Parameters  &   Units           &   Value (error, 90\%\ confidence) \\
\hline
tbabs            & \nh\          &   10$^{22}$ \cms\    & 0.25 (+0.05, -0.07)\\
blackbody        & kT            &   eV               & 5.25 (frozen) \\
                 & Normalisation &                    & 2.49 (+0.03, -0.05)\\
phabs            & \nh\          &   10$^{22}$ \cms\    & tied to that of the tbabs\\
Part. Cvr. Abs.  & \nh\          &   10$^{22}$  \cms\   & 4.6 (+0.3, -0.4)\\
                 & Cvf. Frc      &                    & 0.951 (+0.012, -0.001)\\
Part. Cvr. Abs.  & \nh\          &   10$^{22}$  \cms\   & 48 (+4, -13)\\
                 & Cvf. Frc      &                    & 0.76 (+0.02, -0.03)\\
blackbody        & kT            &   eV               & 57 (+3, -2) \\
                 & Normalisation &                    & 0.43 (+0.17, -0.09) \\
`Cropper'        & \Mdot         &   g s$^{-2}$ \cms  & 0.51 (+18.1, -0.05)\\
                 & Normalisation &                    & 1.10\tim{-4} (+18.9, -2.7\tim{-5})\\
\hline
\end{tabular}
\caption{The best-fitting parameters for the \swift-UVOT and XRT
spectrum of observation 009.}
\label{tab:spec}
\end{center}
\end{table*}

\subsection{Spin-period modulation} 
\label{sec:spin}

\begin{figure*}
\begin{center}
\hbox{
\psfig{file=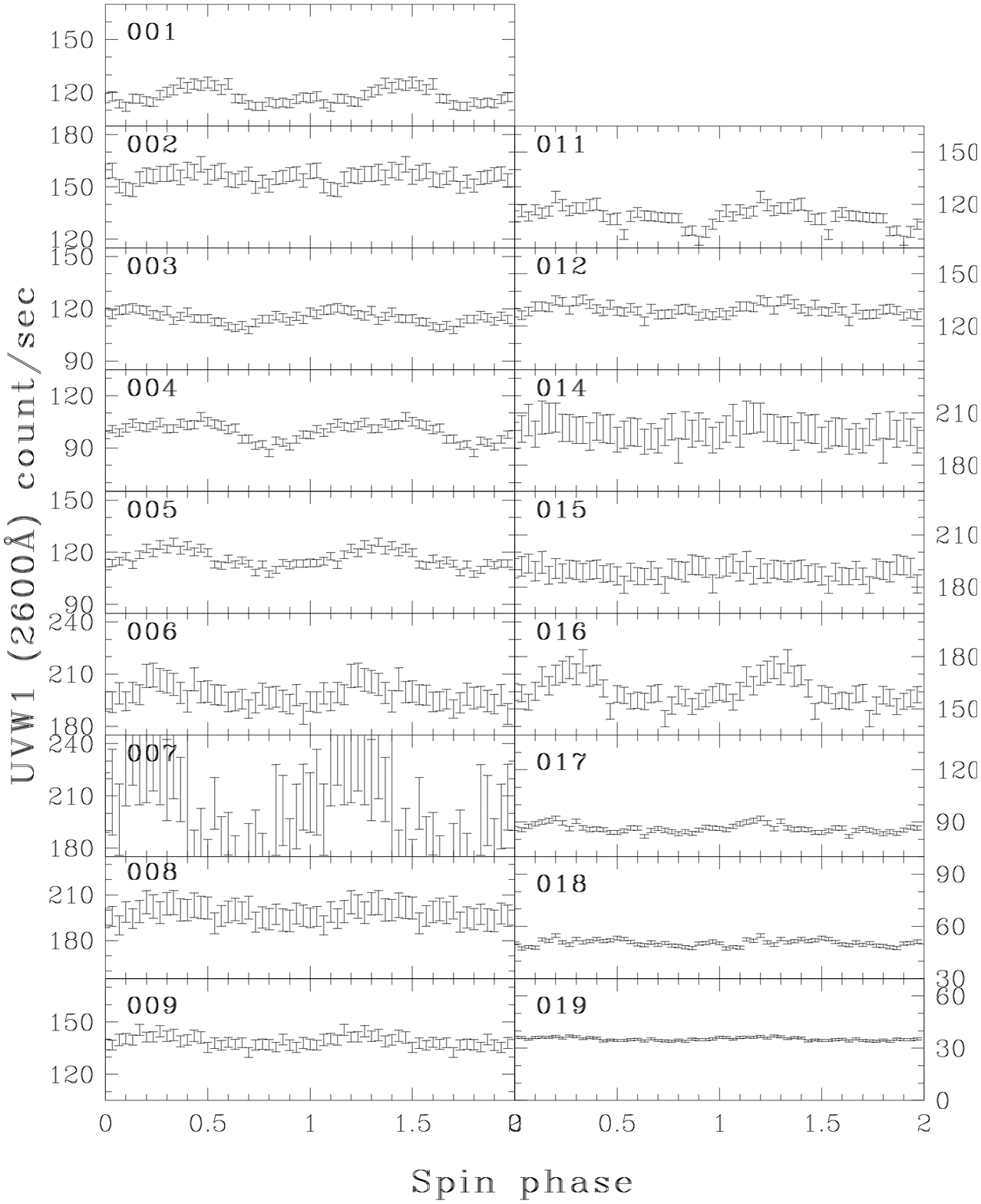,width=8.1cm}
\hspace{1cm}
\psfig{file=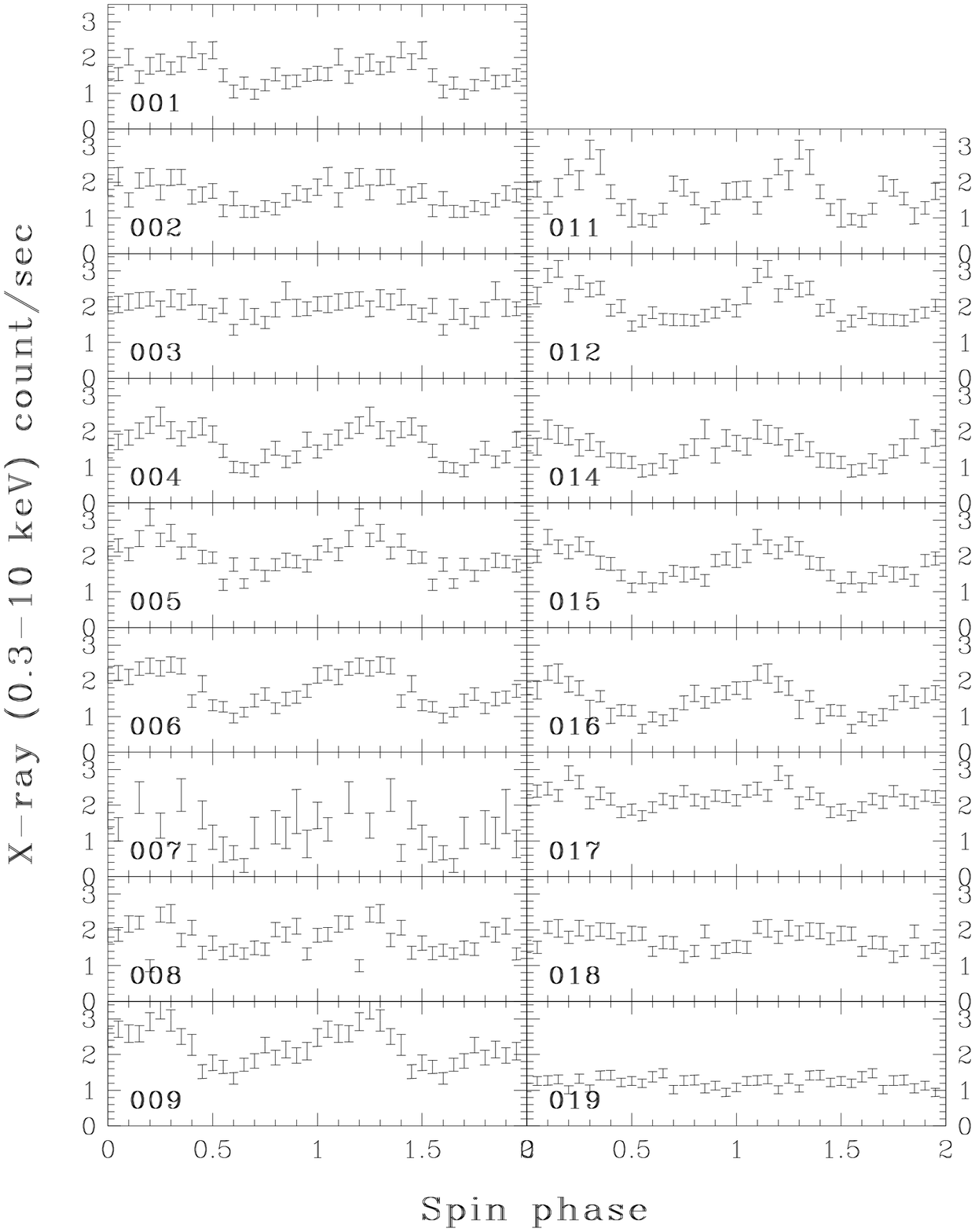,width=8.1cm}
}
\caption{UV (2600\AA) and X-ray (0.3-10 keV) spin-folded light curves of
GK Per. All of the plots have the same (arbitrary) zero-point in the
barycentric reference-frame. Note that the UV panels have different
$y$-axes since the emission was so variable, however each has a range of
70 \cps.}
\label{fig:spin}
\end{center}
\end{figure*}

One of the signatures of IPs is that their emission is modulated on the
white-dwarf spin period. For GK Per this is 351 s (Watson \etal1985;
Mauche 2004). We folded the X-ray and UV data for each observation on
this period -- using the same arbitrary zero-point (T0+52.65 s, in the
barycentric frame) each time. The resultant folds are shown in
Fig.~\ref{fig:spin}. 

As can be seen the shape, magnitude and phase of spin minimum changes
from observation to observation. This behaviour has been  seen in
quiescence, both in X-rays (Norton \etal1988) and in the $V/R$ ratios of the
H$\alpha$ and H$\delta$ Balmer lines (Garlick \etal1994).

In some cases the shape of the profile is the same in both wavebands and
in others they differ (for example, observation 001 has a roughly
sawtoothed profile in each band whereas the UV profile in observation
009 is much more symmetric than the X-ray profile).

Unfortunately the individual observations contain too few counts  for a
meaningful phase-resolved spectroscopic analysis. We are reluctant to
combine observations for this purpose because of the pulse-profile
evolution. Instead we created hardness ratios and folded these on the
spin period. Because the source is so heavily absorbed it was necessary
to use the 4--10/0.3--4 keV hardness ratio in order to have sufficient
counts in the `soft' band, however even with this ratio the rate in that
band is so low that large bins and hence low time resolution is
necessary. The hardness ratio spin folds are shown in
Fig.~\ref{fig:hardspin}. Little significant modulation is seen, however
this is not entirely surprising: the spin-period modulation is thought
to be an absorption effect (Hellier \etal2004; Vrielmann \etal2005), and
our hardness ratio is not especially sensitive to absorption.

A \til5000-s quasi-periodic oscillation (QPO) has been previously
reported in GK Per outburst observations (e.g.\ Watson \etal1985;
Hellier \etal2004). Unfortunately, the orbital period of \swift\ is
close to this; we thus do not consider the QPO further in this paper.

\section{Discussion}
\label{sec:disc}

\subsection{Interpreting the outburst profile}
\label{sec:discoutburst}

We have presented a high-quality multi-wavelength dataset  covering the
2006 outburst of GK Per, which shows that this was an atypical event.
Lasting 20--30 days longer than a typical outburst, the optical
brightness peaked 1.5 magnitudes below that seen in most previous
events. The \til2600 \AA\ UV flux is similarly reduced compared to
previous outbursts. In contrast, the hard X-ray flux is consistent with
that seen in previous outbursts.

At first glance these statements seem paradoxical. The optical flux
tracks the disc brightness. This brightness indicates the extent of the
region of the disc which is in outburst. Thus the lower luminosity seen
in 2006 suggests that less of the disc was in outburst than in previous
years, hence less mass was transferred through the disc. This is further
supported by Fig.~\ref{fig:fluence}, which showed the optical fluence of
the 2006 outburst to be abnormally low. In contrast, the hard X-ray flux
in IPs is often assumed to track the rate of accretion onto the white
dwarf. The typical outburst X-ray flux seen in 2006, combined with the
long duration of this outburst, therefore implies that more mass was
accreted in 2006 than during typical outbursts. These inferences cannot
both be true.

A resolution of this apparent inconsistency lies in the joint X-ray and
UV spectral fit (Section~\ref{sec:spec}). This revealed that the
proportion of the accretion flux emitted in the 2--10 keV band could be
as little as 1\% of the total radiated accretion flux: the flux in this
energy range is clearly not a good proxy for the accretion rate.

In order to understand the unusual nature of the 2006 outburst we must
consider the accretion disc, since DN outbursts are thought to be
disc-instability events (e.g.\ Lasota 2001). GK~Per is an atypical DN as
it is an IP, thus the inner disc is missing. Although this may have an
effect on the shape and duration of outbursts in GK~Per compared to
other DNe, it is unlikely to be the cause of the unusual nature of the
2006 outburst, since the magnetic field of the white dwarf should affect
all outbursts in a similar way. GK~Per is also atypical due to its long
(2-day) orbital period, meaning the system contains a much larger disc
than most CVs. Warner (1995) gives the outer disc radius as

\begin{equation}
R_d=\frac{0.6}{1+q}
\label{eq:rd}
\end{equation}

which for GK Per evaluates to $R_d=2\tim{11}$ cm, assuming  the
masses of the white dwarf and secondary to be 0.87 \msol\ and 0.48
\msol\ respectively (Morales-Rueda \etal2002). The viscous timescale of
the disc will thus also be much longer than is typical for CVs. The
viscous timescale at radius $R$ is given by 

\begin{figure}
\begin{center}
\psfig{file=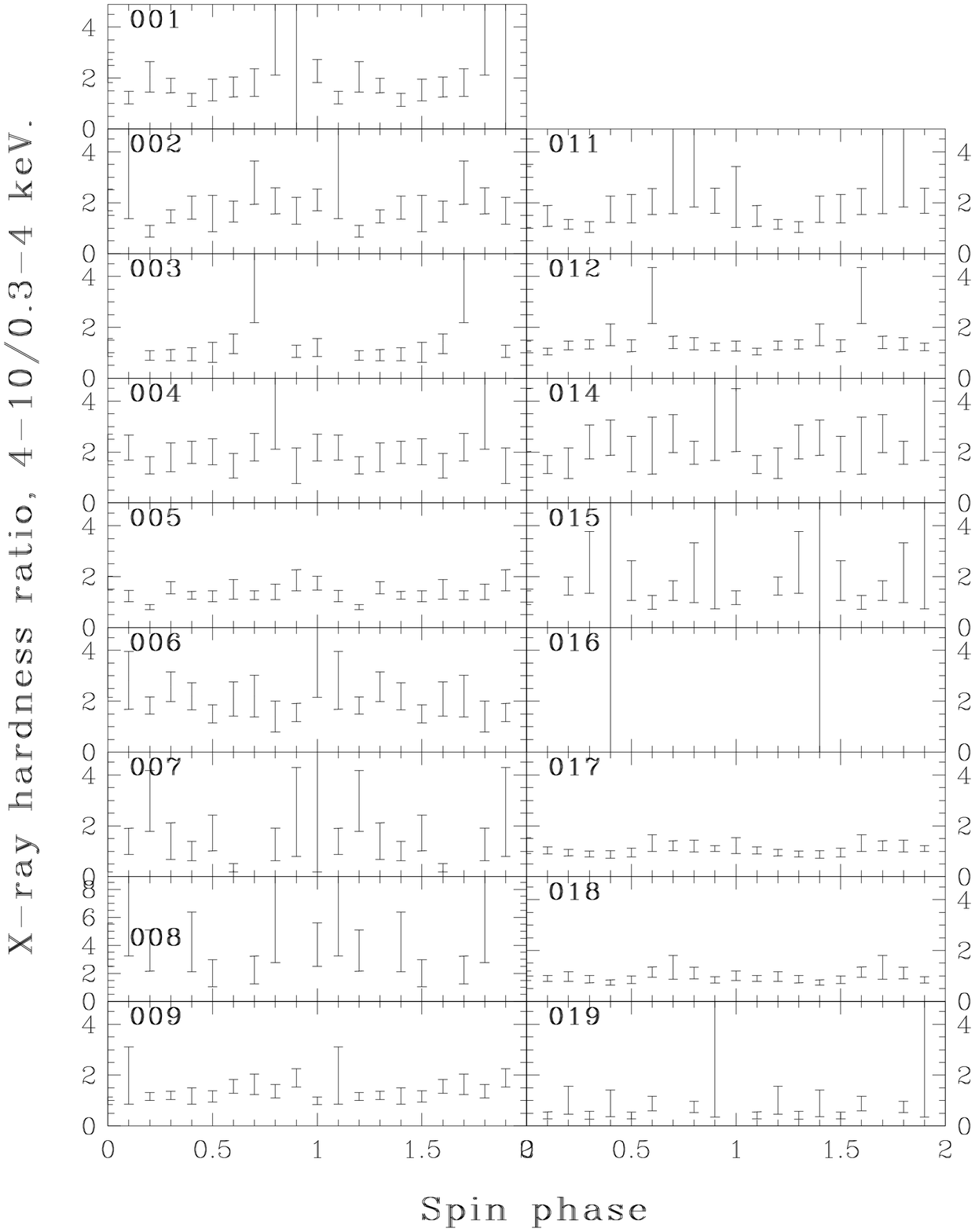,width=8.1cm}
\caption{X-ray hardness ratios (4--10/0.3--4 keV) of GK Per folded on the
351-s spin period. The zero point is the same as for
Fig.~\ref{fig:spin}.}
\label{fig:hardspin}
\end{center}
\end{figure}

\begin{equation}
\tau_v = \frac{R^2}{\alpha c_s H}
\label{eq:tvisc}
\end{equation}

where $\alpha$ is the dimensionless viscosity parameter ($\alpha$\til0.1
during outburst and at least a factor of 10 lower in quiescence), $c_s$
is the sound speed in the disc, (\til$10^6$ cm s$^{-1}$) and $H$ is the
scale height of the disc (\til$0.05R$). The cold-state (i.e. quiescent)
viscous timescale at the outer edge of the disc is thus
$\tau_{q,R_d}$\til4600 days, which is longer than the typical
inter-outburst time of \til1100 days. We also note that this timescale
is only a factor \til3 different from the delay between the 1967
outburst and the 2006 outburst, i.e.\ the large disc is capable of
producing variations on the approximate timescale on which the outburst
morphology is showing variation.

A typical outburst lasts \til70 days (Fig.~\ref{fig:fluence}).
Interpreting this as a viscous decay timescale, inverting
equation~(\ref{eq:tvisc}) shows that such outbursts extend to a disc
radius $R$\til3\tim{10} cm, i.e. about 10\%\ of the disc (by radius) is
involved in a typical outburst.

The above numbers demonstrate that the disc in GK Per can retain a
`memory' of its state which is not erased either by outbursts or during
the quiescent inter-outburst period. This is because the cold-state
viscous timescale  is longer than the inter-outburst interval for a
significant fraction of the disc. Thus if there were, for example,
long-term variations in the mass transfer rate from the secondary (e.g.\
caused by magnetic activity on the star), these would be reflected in
the disc density profile for many years. Further, the disc configuration
before and after any given outburst will vary. The fact that the 2006
outburst was different from previous events is thus not surprising: it
is entirely possible that long term changes in the mass transfer rate
are embedded in the disc density profile and hence outburst light
curves. That many outbursts are similar to each other (and even the 2006
outburst follows a `typical' outburst pattern for the first 10--15 days)
is still consistent with this idea: the outburst is triggered when the
surface density somewhere in the disc reaches the (radius-dependent)
critical value. If each outburst begins at around the same place then by
definition the disc state at this point must be approximately the same
at the start of each outburst, hence the outbursts will appear similar
at early times. As the heating wave propagates outwards to radii where
the disc state can differ from outburst to outburst it becomes possible
to observe variation in the outburst profile.

In general terms, this idea allows for long-term variations in the
outburst profile, we consider now the detailed shape of the 2006
outburst and how this can be explained. 

The shape of the optical and UV outburst light curve
(Fig.~\ref{fig:outburst}) is suggestive of 3 short, faint outbursts
running into each other (each reminiscent of the 1970 outburst, for
example). A possible interpretation is thus that a series of heating and
cooling waves passed through the disc giving a mini-outburst which is
twice rekindled. This could be achieved if the heating
wave is triggered at the inner disc but encounters a lower-density which
halts its progress. At this point, as is usual for the end out DN
outbursts, a cooling wave is launched (Lasota 2001). This wave travels
inwards, reducing the amount of the disc which is in the hot state.
However this short time is less than the viscous timescale at the
outburst triggering radius, so the outburst in the inner regions of the
disc is not extinguished. The cooling wave is reflected back as a
heating wave and the outburst is rekindled. We suggest that this
sequence of events happens twice, giving rise to the triple-peaked
optical/UV outburst profile.

If this idea is correct, i.e\ the outburst profile is determined by the
disc density structure, then the similarity between the 1967 and 2006
outbursts suggests that the next outburst will be shorter and less
luminous than normal, akin to the 1970 outburst.

\subsection{The spin-period modulation}
\label{sec:discspin}

The origin of the X-ray spin-period pulsations in GK Per was discussed
extensively by Hellier \etal(2004) and Vrielmann \etal(2005). They
proposed that the modulation is caused by varying absorption as the
`accretion curtains' of magnetically confined material pass through our
line-of-sight to the emitting regions, although the specific geometric
details differ between those two papers. If this is correct we would
expect the hardness ratio to show a maximum of hardness at spin minimum
(i.e.\ when absorption is at its greatest).
Figs.~\ref{fig:spin} and~\ref{fig:hardspin} appear to support this, although
the errors on the hardness ratio are too large to make a definitive
statement. The \xmm\ spin-pulse profile and hardness ratio from the 2002
outburst (Evans \&\ Hellier 2005) however shows this correlation
clearly.

The phasing and shape of the X-ray (and UV) spin-period modulation
varies during the outburst (Fig.~\ref{fig:spin}). Norton \etal(1988)
reported a similar effect in quiescent X-ray data. They noted that in
quiescence the accretion rate is barely enough to overcome the
magnetospheric boundary and produce stable accretion, so the
accretion may be time-dependent and thus the accretion geometry will be
variable, explaining the changing pulse profiles. Clearly in outburst
the accretion rate is much higher and away from this limit, however the
multi-wavelength light curves (Fig.~\ref{fig:outburst}) show a
significant amount of variability from observation to observation,
suggesting that the accretion rate is not stable. This in turn means
that the disc--magnetosphere interaction region will be constantly
changing, thus an unvarying spin-pulse profile is not expected.

\section{Conclusions}

We have presented a unique, multi-wavelength, high-cadence dataset
monitoring the evolution of the 2006 outburst of GK Per at optical, UV
and X-ray wavelengths. The optical outburst profile is unusual, showing
three weak peaks, rather than the typical single, bright peak; it is
also \til30\%\ longer than a typical outburst. The UV data follow the
optical evolution. The X-ray data, in contrast, appear entirely
consistent with previous outbursts. This presents a significant
challenge to existing disc outburst models.

We have shown that the large disc in GK~Per is able to maintain a
long-term `memory' of its state (and hence the mass transfer rate from
the secondary) which is not erased by outbursts or by quiescent
accretion during the inter-outburst period. This is expected to produce
long-term variation in outburst morphology. Within this context we
interpret the 2006 outburst as a short outburst which is thrice
suppressed by low density regions of the disc, and twice rekindled by
the high density inner regions. We also suggest that the next outburst,
expected around 2009--2010, will be shorted than normal, similar to the
1970 outburst.

\section*{Acknowledgements}

We acknowledge with thanks the variable star observations from the AAVSO
International Database contributed by observers worldwide and used in
this research. We thank the \swift\ PI, Neil Gehrels, and the \swift\
science planners for supporting the ToO observations. This work made use
of data supplied by the UK Swift Science Data Centre at the University
of Leicester. PAE, APB, JPO acknowledge STFC support.

\label{lastpage}
\end{document}